\documentclass[twocolumn]{article}

\usepackage{msanii}         % styling

\usepackage[utf8]{inputenc} % allow utf-8 input
\usepackage[english]{babel} % language support
\usepackage{url}            % simple URL typesetting
\usepackage{booktabs}       % professional-quality tables
\usepackage{amsfonts}       % blackboard math symbols
\usepackage{amsmath}        % math
\usepackage{nicefrac}       % compact symbols for 1/2, etc.
\usepackage{microtype}      % micro-typography
\usepackage{cleveref}       % smart cross-referencing
\usepackage{graphicx}       % graphics management
\usepackage{natbib}         % bibliography management
\usepackage{abstract}       % control abstract typesetting
\usepackage{float}          % control figure placement

\title{Msanii: High Fidelity Music Synthesis on a Shoestring Budget}

\author{
    Kinyugo Maina \\
    Independent Researcher \\
    \texttt{kinyugomaina@gmail.com}
}

\begin{document}

\twocolumn[
    \maketitle % need full-width title
    \begin{onecolabstract}
        \noindent
        \begin{quote}
            In this paper, we present Msanii, a novel diffusion-based model for synthesizing long-context, high-fidelity music efficiently. Our model combines the expressiveness of mel spectrograms, the generative capabilities of diffusion models, and the vocoding capabilities of neural vocoders. We demonstrate the effectiveness of Msanii by synthesizing tens of seconds (\textit{190 seconds}) of \textit{stereo} music at high sample rates (\textit{44.1 kHz}) without the use of concatenative synthesis, cascading architectures, or compression techniques. To the best of our knowledge, this is the first work to successfully employ a diffusion-based model for synthesizing such long music samples at high sample rates. Our demo can be found \href{https://kinyugo.github.io/msanii-demo}{here} and our code \href{https://github.com/Kinyugo/msanii}{here}.
        \end{quote}
        \vspace{0.5in}
    \end{onecolabstract}
]

\section{Introduction}
Music is a universal language that elicits emotions and connects people from diverse cultures, and is an integral part of society. For decades, researchers have been investigating whether computers can capture the creative process behind music creation and the potential implications for music and artificial intelligence.

In recent years, the field of generative modeling has seen significant growth with various techniques, including generative adversarial networks (GANs) \cite{goodfellow2020generative}, variational autoencoders (VAEs) \cite{kingma2013auto}, normalizing flows \cite{rezende2015variational}, autoregressive models \cite{dhariwal2020jukebox}, and diffusion models \cite{sohl2015deep,ho2020denoising,kingma2021variational}, driving progress in various fields. These techniques have achieved human-level performance in tasks such as image generation \cite{rombach2022high,karras2020analyzing,dhariwal2021diffusion,saharia2022photorealistic,ramesh2022hierarchical}, speech generation \cite{kong2020diffwave,shen2017natural}, and text generation \cite{brown2020language,scao2022bloom}, as well as progressed music generation \cite{dhariwal2020jukebox,rouard2021crash,engel2019gansynth,marafioti2019adversarial} and other areas.

However, efficient high-fidelity music synthesis remains a challenging task in machine learning due to the high dimensionality of audio signals, which makes it difficult for models to learn the long-range structure of music \cite{dhariwal2020jukebox,dieleman2018challenge}. To address this issue, it is common to learn a lower-dimensional representation of the audio signal, which can reduce computational complexity and allow models to better capture the salient features of music related to fidelity \cite{dhariwal2020jukebox, dieleman2018challenge}. As an alternative to learning a lower-dimensional representation, Time-Frequency (TF) representations, such as mel spectrograms, provide a powerful and intuitive way to represent features in audio signals. Mel spectrograms, which are a type of TF feature, have been widely used in various applications, including natural language processing \cite{shen2017natural}, voice conversion \cite{hwang2020mel}, and singing voice synthesis \cite{liu2022diffsinger}. They have also been applied successfully to music synthesis \cite{engel2019gansynth,marafioti2019adversarial,pasini2022musika}. Mel spectrograms are particularly appealing for music synthesis due to their low resolution, which allows them to capture important musical characteristics while minimizing computational complexity.

Autoregressive models and GANs are popular choices for music synthesis, but they each have their own challenges. Autoregressive models, which have been widely used in the raw waveform domain \cite{dhariwal2020jukebox, dieleman2018challenge}, are often slow at inference. GANs, which have frequently been employed for music synthesis using TF representations \cite{engel2019gansynth, marafioti2019adversarial, pasini2022musika}, can suffer from unstable training and the adversarial training of multiple networks, leads to low sample diversity, in addition to being computationally expensive. In contrast, diffusion-based models offer fast inference compared to autoregressive models, a simple training procedure, and have recently outperformed GANs in terms of quality \cite{dhariwal2021diffusion}. This makes diffusion-based models an attractive choice for music synthesis.

In this paper, we propose a novel approach for music synthesis using mel spectrograms that leverages the benefits of diffusion-based modeling. By combining the expressiveness of mel spectrograms with our novel U-Net architecture and diffusion models, we are able to synthesize minutes of high-fidelity music at a high sample rate. Our method represents a significant advance in the field of music synthesis, as it generates long samples of high-quality music without relying on concatenative synthesis, cascading architectures, or compression techniques. Additionally, we show that our model, Msanii, can be used to solve other audio tasks, such as audio inpainting and style transfer, without the need for retraining.

The main contributions of this paper are:
\begin{itemize}
    \item We introduce Msanii, a novel diffusion-based model for long-context, high-fidelity music synthesis in the mel spectrogram domain. To the best of our knowledge, this is the first work to successfully employ a diffusion-based model for synthesis of minutes of audio at high sample rates (44.1 kHz) in the TF domain.
    \item We demonstrate the effectiveness of Msanii by synthesizing tens of seconds (\textit{190 seconds}) of \textit{stereo} music at a high sample rate (\textit{44.1 kHz}).
    \item We show that Msanii can be used to solve other audio tasks, such as interpolation, style transfer, inpainting and outpainting, without the need for retraining.
\end{itemize}

The rest of the paper is organized as follows: in Section \ref{sec:background} we review related work in music synthesis and diffusion-based models. In Section \ref{sec:architecture} we describe our proposed method, including the architecture and training of Msanii. In Section \ref{sec:experiments} we present our experimental setup. In Section \ref{sec:results} we present our results. In Section \ref{sec:futurework} we discuss potential future work. And Finally, in Section \ref{sec:conclusion} we summarize our contributions.

\textbf{DISCLAIMER:} This paper is a work in progress and has not been finalized. The results presented in this paper are subject to change and should not be considered final.

\section{Background}
\label{sec:background}

The high dimensionality of audio signals presents a significant challenge for music synthesis in the raw waveform domain. To accurately represent an audio sample, it is necessary to discretize the continuous signal into a large number of samples, which requires a high sample rate (thousands of samples per second). For music at CD quality, this sample rate is typically 44.1 kHz. This means that a ${\sim}3$ minute long audio sample will consist of approximately ${\sim}8$ million samples. This complexity increases with the number of channels, as the total number of samples $T$ becomes $T = duration \times sample\ rate \times channels$.

The computational demands of synthesizing long audio samples are further compounded by the need to capture a wide range of musical structures, such as timbre, harmony, and melody, as well as to ensure global coherence in terms of form and texture. To address these challenges, it is common to use a lower-dimensional representation of the audio signal that captures important musical features while minimizing computational complexity. Additionally, it is necessary to employ an expressive but efficient generative model.

\subsection{Mel Spectrograms}
To address the computational complexity of our task, we propose the use of a lower-dimensional yet expressive representation of audio: Mel Spectrograms. Mel spectrograms are a popular representation of audio used in tasks such as speech synthesis \cite{shen2017natural}, voice conversion \cite{hwang2020mel}, and music synthesis \cite{vasquez2019melnet}. They are derived from the magnitude spectrogram of the Short-Time Fourier Transform (STFT) and encode frequency in a way that is more perceptually relevant to human hearing. However, the conversion from raw audio to mel spectrograms is not perfectly invertible due to the loss of phase information in the magnitude STFT spectrogram.

To reconstruct audio from mel spectrograms, neural vocoders such as MelGAN \cite{kumar2019melgan}, ISTFTNet \cite{kaneko2022istftnet}, MCNN \cite{arik2018fast} and Phase Gradient \cite{di2022mel} have been developed to approximate both the magnitude and phase from the mel spectrogram. However, designing a vocoder that reconstructs both the magnitude and phase while remaining lightweight can be challenging. As an alternative, we propose reconstructing only the magnitude spectrogram and approximating the phase using traditional methods. This approach has been demonstrated in previous work such as Adversarial Audio Synthesis \cite{marafioti2019adversarial} and MelNet \cite{vasquez2019melnet}. In particular, we use the Griffin-Lim algorithm \cite{griffin1984signal, perraudin2013fast}.

To reconstruct the magnitude spectrogram, we use a combination of the Spectral Convergence Loss and Log-Magnitude Loss. The Spectral Convergence Loss is defined as:

\begin{align}
    \frac{\||STFT(s)| - |STFT(s)|\|_{F}}{\||STFT(s)|\|_{F}}
\end{align}

where $\left| . \right|$ represents the magnitude, $| . |_{F}$ is the Frobenius norm, and $s$ and $\hat{s}$ are the ground truth and predicted magnitude spectrograms, respectively. This loss focuses on the large magnitude components of the spectrograms.

The Log-Magnitude Loss is defined as:

\begin{align}
    \|log(|STFT(s)| + \epsilon) - log(|STFT(\hat{s})| + \epsilon)\|_{1}
\end{align}

where $\epsilon$ is a small constant value added to prevent taking the logarithm of zero, and $| . |_{1}$ is the $L1$ norm. This loss focuses on the small magnitude components of the spectrograms.

\subsection{Diffusion}
Inspired by the recent successes of diffusion models \cite{sohl2015deep,ho2020denoising,kingma2021variational} in solving audio tasks \cite{rouard2021crash,kong2020diffwave}, we chose to employ them for the task of synthesizing mel spectrograms. Diffusion models can be thought of as a Markovian Hierarchical Variational Autoencoder \cite{luo2022understanding}. They define a markov chain of steps to slowly add random noise to the data and then learn the reverse process to synthesize data samples from noise.

\subsubsection{Forward Process}
Given a real data distribution $q(\vx_0)$, we draw a sample $\vx_0$ from the distribution, $\vx_0 \sim q(\vx)$. Then, we define a forward noising process $q(\vx_t|\vx_{t-1})$ that gradually adds Gaussian noise to the sample according to a predefined schedule ${\beta_t \in (0,1)}_{t=1}^{T}$, where $\beta_{1} < \beta_{2} < \dots < \beta_{T}$. Specifically, the sample distribution at each time step is given by:

\begin{align}
    q(\vx_t|\vx_{t-1}) & = \mathcal{N}(\vx_t;\sqrt{1 - \beta_t}\vx_{t-1},\beta_t\idty)
\end{align}
And the distribution over the entire sequence of samples $\vx_{1:T}$ given the initial sample $\vx_0$ is given by:

\begin{align}
    q(\vx_{1:T}|\vx_0) & = \prod_{t = 1}^{T}q(\vx_{t}|\vx_{t-1})
\end{align}
As $T \rightarrow \infty$, the distribution of $\vx_T$ approaches the standard Gaussian distribution.

To sample from the forward distribution, we can draw a sample $\vx_{t}$ at each time step $t$ from a conditional Gaussian with mean $\mu_t=\sqrt{1-\beta_t}\vx_{t-1}$ and variance $\sigma^2=\beta_t$ as follows:
\begin{align}
    \vx_t = \sqrt{1-\beta_t}\vx_{t-1} + \sqrt{\beta_t}\beps & \quad \text{where } \beps \sim \mathcal{N}(\vzero,\idty)
\end{align}

The forward noising process in diffusion models has the property that, using the reparameterization trick, we can sample $\vx_t$ at any arbitrary timestep. This property is described in \cite{song2020denoising,luo2022understanding,weng2021diffusion}. Let $\alpha_t = 1 - \beta_t$ and $\bar{\alpha}_t = \prod_{i=1}^t \alpha_i$. The process can be expressed as follows:

\begin{align}
    \label{eqn:forwardnoisingprocess}
    \begin{split}
        \vx_t &= \sqrt{\alpha_t}\vx_{t-1} + \sqrt{1-\alpha_t}\beps_{t-1} \\
        &=\sqrt{\alpha_t\alpha_{t-1}}\vx_{t-2} + \sqrt{1 - \alpha_t\alpha_{t-1}}\bar{\beps}_{t-2}  \\
        &= \dots \\
        &=\sqrt{\alpha_t}\vx_0 + \sqrt{1-\alpha_t}\beps
    \end{split}
\end{align}
where, $\beps_{t-1},\beps_{t-2}, \dots \sim \mathcal{N}(\vzero,\idty)$. $\bar{\beps}_{t-2}$ is a combination of two Gaussian distributions with different variances, $\mathcal{N}(\vzero, \sigma^2_1\idty)$ and $\mathcal{N}(\vzero, \sigma^2_2\idty)$, such that:

\begin{align}
    \bar{\beps}_{t-2}=\sqrt{(1 - \alpha_t) + \alpha_t(1 - \alpha_{t-1})} = \sqrt{1 - \alpha_t\alpha_{t-1}}
\end{align}

\subsubsection{Reverse Process}
\label{sssec:reverseprocess}
To generate a sample from a Gaussian noise input $\vx_T \sim \mathcal{N}(\vzero,\idty)$, we reverse the forward process by sampling from $q(\vx_{t-1}|\vx_t)$. However, $q(\vx_{t-1}|\vx_t)$ depends on the entire dataset, so we learn an approximate model $p_{\theta}$:

\begin{align}
    p_{\theta}(\vx_{t-1}|\vx_t) & = \mathcal{N}(\vx_{t-1};\mu_{\theta}(\vx_t, t), \Sigma_{\theta}(\vx_t, t)) \\
    p_{\theta}(\vx_{0:T})       & = p(\vx_T)\prod^T_{t=1}p_{\theta}(\vx_{t-1}|\vx_t)
\end{align}

The reverse conditional probability becomes tractable when it is conditioned on $\vx_0$:

\begin{align}
    q(\vx_{t-1}|\vx_t,\vx_0) = \mathcal{N}(\vx_{t-1};\tilde{\mu}_t(\vx_t,\vx_0), \tilde{\Sigma}_t(\vx_t,\vx_0))
\end{align}

Similar to \cite{ho2020denoising}, we fix the variance to:
\begin{align}
    \Sigma_{\theta}(\vx_t, \vx_0) = \sigma^2_t\idty = \tilde{\beta}_t\idty
\end{align}
and learn only the mean $\mu_{\theta}$. This gives us:

\begin{align}
    q(\vx_{t-1}|\vx_t,\vx_0) = \mathcal{N}(\vx_{t-1};\tilde{\mu}_t(\vx_t,\vx_0), \tilde{\beta}_t\idty)
\end{align}

Using Bayes' rule, we can derive the following equations:

\begin{align}
    \tilde{\mu}_t(\vx_t,\vx_0) & = \frac{\sqrt{\alpha_t}(1 - \bar{\alpha}_{t-1})}{1 - \bar{\alpha}_t} \vx_t + \frac{\sqrt{\bar{\alpha}_{t-1}\beta_t}}{1 - \bar{\alpha}_{t-1}} \vx_0 \\
    \tilde{\beta}_t            & = \frac{1 - \bar{\alpha}_{t-1}}{1 - \bar{\alpha}_t} \cdot \beta_t
\end{align}

By substituting $\vx_0 = \frac{1}{\sqrt{\bar{\alpha}_t}}(\vx_t - \sqrt{1 - \bar{\alpha}_t}\beps_t)$ into the above equation for $\tilde{\mu}_t(\vx_t,\vx_0)$, we get:

\begin{align}
    \tilde{\mu}_t(\vx_t,\vx_0) = \frac{1}{\sqrt{\bar{\alpha}_t}}(\vx_t - \frac{1 - \alpha_t}{\sqrt{1 - \bar{\alpha}_t}} \beps_t)
\end{align}

This equation allows us to compute the mean of the distribution $q(\vx_{t-1}|\vx_t,\vx_0)$ given $\vx_t$ and $\vx_0$. The variance of this distribution is fixed to $\tilde{\beta}_t\idty$, where $\tilde{\beta}_t$ is given by the equation $\tilde{\beta}_t = \frac{1 - \bar{\alpha}_{t-1}}{1 - \bar{\alpha}_t} \cdot \beta_t$. We can then sample from this distribution to obtain a sample of $\vx_{t-1}$ given $\vx_t$ and $\vx_0$. This process can be repeated to generate samples of $\vx_{t-2}, \vx_{t-3}, \dots, \vx_0$ given $\vx_T$.

\subsubsection{Loss Function}
As $\vx_t$ is available during training, we can use it to reparameterize the Gaussian mean term and predict $\beps_t$ given $\vx_t$ and $t$. This is done with the following equation:

\begin{align}
    \tilde{\mu}_t(\vx_t,\vx_0) = \frac{1}{\sqrt{\bar{\alpha}_t}}(\vx_t - \frac{1 - \alpha_t}{\sqrt{1 - \bar{\alpha}t}} \beps_{\theta}(\vx_t, t))
\end{align}

We then use the simplified loss term as described by \cite{ho2020denoising}. Resulting in the following equations:

\begin{align}
    \begin{split}
        L_t & = \mathbb{E}_{t \sim [1,T],\vx_0,\beps_t}[\|\beps_t - \beps_{\theta}(\vx_t, t)\|^2_2]                                           \\
        & = \mathbb{E}_{t \sim [1,T],\vx_0,\beps_t}[\|\beps_t - \beps_{\theta}(\sqrt{\alpha_t}\vx_0 + \sqrt{1-\alpha_t}\beps_t, t)\|^2_2]
    \end{split}
\end{align}

\section{Architecture}
\label{sec:architecture}
\begin{figure*}
    \centering
    \includegraphics[scale=0.45]{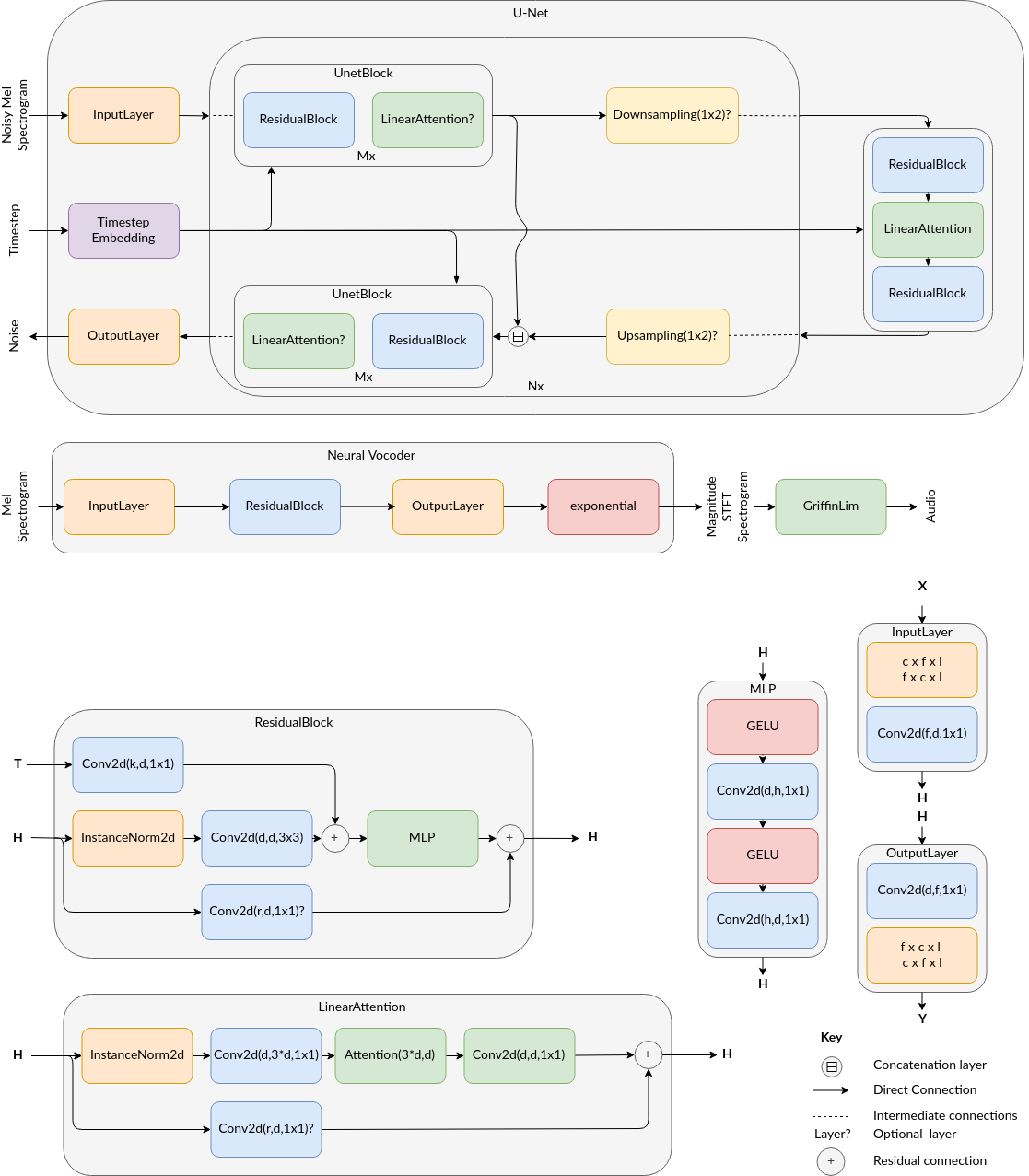}
    \caption{Proposed Msanii Architecture.}
    \label{fig:msaniiarchitecture}
\end{figure*}

Motivated by the success of the patch embedding tokenization scheme in the Vision Transformer (VIT) \cite{dosovitskiy2020image}, we propose a similar tokenization scheme for audio based on mel spectrograms. Specifically, we view mel spectrograms as a sequence of tokens, where the sequence length is equal to the time frames and the dimensionality of each token is equal to the number of mel frequencies. This is similar to taking a patch along the frequency dimension, with a patch size equal to the number of mel frequencies.

In contrast to other methods that treat mel spectrograms as images, our approach offers efficiency gains by reducing the context size. For a mel spectrogram with dimensions $channels \times frequencies \times time\ frames$, our context size is reduced to $channels \times time\ frames$. Furthermore, we process channels independently, allowing our method to be applicable to an arbitrary number of channels. These considerations have been incorporated into the design of our U-Net (see Section \ref{ssec:unet}) and Neural Vocoder (see Section \ref{ssec:neuralvocoder}).

\subsection{U-Net}
\label{ssec:unet}

U-Nets \cite{ronneberger2015u} are a popular choice for image segmentation tasks due to their ability to accurately model fine details and localize features through the use of skip connections. These neural networks have also been greatly applied to diffusion modeling \cite{nichol2021improved,song2020denoising,ho2020denoising}. In this work, we propose a U-Net architecture that combines the strengths of U-Nets with those of transformers \cite{vaswani2017attention}, which are known for their ability to capture long-range dependencies through self-attention mechanisms. The resulting model is able to capture both local and global context while remaining efficient (see Figure \ref{fig:msaniiarchitecture}). We present components of the U-Net below.

\subsubsection{Input and Output Layers}
\label{sssec:inputandoutputlayers}

The input layer (tokenization layer) receives a 2D spectrogram $\mathbf{X} \in \mathbb{R}^{c \times f \times l}$, where $c$ is the channel dimension (e.g., mono, stereo), $f$ is the frequency dimension, and $l$ is the time frames dimension. It reshapes and linearly projects it along the frequency dimension to obtain $\mathbf{H} \in \mathbb{R}^{d \times c \times t}$, where $\mathbf{H}$ is the output, $d$ is the frequency dimension after projection.

The output layer (detokenization layer) expects an input $\mathbf{H} \in \mathbb{R}^{d \times c \times l}$ and performs the inverse actions of the input layer, linearly projecting the input and reshaping it into $\mathbf{Y} \in \mathbb{R}^{c \times f \times l}$, where $\mathbf{Y}$ is the output (see Figure \ref{fig:msaniiarchitecture}).

\subsubsection{Residual Block}
\label{sssec:residualblock}

The input to our residual block is a 2D latent feature $\mathbf{H} \in \mathbb{R}^{d \times c \times l}$ and a timestep embedding feature $\mathbf{T} \in \mathbb{R} ^ {k \times 1 \times 1}$, where $d$ is the frequency dimension, $c$ is the channel dimension, $l$ is the time frames dimension and $k$ is the timestep. We first apply a pre-normalization layer to the input features, then perform a $3 \times 3$ padded convolution without changing the number of features to ensure that the input and output features have the same dimensions. While the U-Net is not sensitive to the specific type of normalization used, we have found that feature normalization performs better than weight normalization. This may be because the use of feature normalization forces the pre-activations to be Gaussian, resulting in output that is also Gaussian, which is useful for modeling noise that is typically Gaussian. We chose to use Instance Norm \cite{ulyanov2016instance} in our implementation, as it has been observed to produce audio with fewer metallic artifacts in our Neural Vocoder \ref{ssec:neuralvocoder}.

After normalizing and projecting the input features, we project the timestep embedding features using a $1 \times 1$ convolution to match the feature dimensions of the latent features. We then sum the timestep embedding features and latent features and feed them into a Multi-Layer Perceptron (MLP). Our MLP is based on ConvNext \cite{liu2022convnet} with a minor modification: we apply the GELU non-linearity before the first convolution layer. Finally, we apply the residual connection to the outputs of the MLP, and optionally project the residual features to match the dimensions of the MLP outputs.

\subsubsection{Linear Attention}
The input to our linear attention block is a 2D latent feature tensor $\mathbf{H} \in \mathbb{R}^{d \times c \times l}$. We first apply Instance Normalization to the input tensor before performing the attention operations. While the choice of attention mechanism does not significantly affect the performance of the U-Net, we have found that Linear Attention \cite{katharopoulos2020transformers} performs well due to its linear computational complexity.

Unlike some transformers \cite{vaswani2017attention,dosovitskiy2020image,katharopoulos2020transformers,liu2022convnet}, we have observed that placing the attention layer after the residual block leads to better global coherence. Therefore, we adopt the post-attention mechanism for all of our tasks. Note that the U-Net is not sensitive to the specific attention mechanism used, which may be due to the fact that attention layers are used in the deeper layers where the context size is already small enough for global context to be easily captured.

\subsubsection{Downsampling and Upsampling Layers}
For downsampling the input features, we use a single $1 \times 3$ convolution layer with a stride of $1 \times 2$ and padding of $0 \times 1$. This effectively reduces the time dimension of the input features by a factor of 2 while preserving the other dimensions.

For upsampling, we use a single $1 \times 4$ transposed convolution layer with a stride of $1 \times 2$ and padding of $0 \times 1$. This effectively increases the time dimension of the input features by a factor of 2 while preserving the other dimensions.

\subsection{Neural Vocoder}
\label{ssec:neuralvocoder}

Our neural vocoder design is inspired by ISTFTNet \cite{kaneko2022istftnet} and features a simple structure. It takes in an input mel spectrogram $\mathbf{X} \in \mathbb{R}^{c \times f_m \times l}$, where $f_m$ is the frequency dimension of the mel spectrogram, and passes it through an input layer (see Section \ref{sssec:inputandoutputlayers}). The mel spectrogram is then processed through a single residual block (see Section \ref{sssec:residualblock}), without the timestep embedding features, and finally through an output layer (see Section \ref{sssec:inputandoutputlayers}). This produces an STFT spectrogram $\mathbf{Y} \in \mathbb{R}^{c \times f_s \times l}$, where $f_s$ is the frequency dimension of the magnitude STFT spectrogram. After the output layer, we apply an exponential activation to transform the magnitude spectrogram from log-space to linear-space (see Figure \ref{fig:msaniiarchitecture}).

\section{Experiments}
\label{sec:experiments}

\subsection{Dataset}

The dataset used for this work is the POP909 dataset \cite{pop909-ismir2020}, which consists of 909 MIDI files of popular pop songs. We synthesize 44.1kHz, stereo audio from the MIDI files using FluidSynth \cite{newmarch2017fluidsynth}.

\subsection{Training Details}

\subsubsection{Data Preprocessing}
To synthesize mel spectrograms from the raw audio, we use an STFT window size of 2048 with a hop size of 1024, and apply 128 mel filterbanks to the resulting mel spectrogram. Since diffusion models typically operate on data in the range $\left[-1,1\right]$ and expect the data to be Gaussian, we propose learning data-specific preprocessing techniques such as moving average parameters for standard scaling (see Algorithm \ref{algo:standardscaling}) and min-max scaling (see Section \ref{appendix:datapreprocessing} for more details).

\subsubsection{Model Training}
For our U-Net model, we set the width to 256 and use 2 U-Net blocks per resolution before applying downsampling/upsampling layers. This results in a total of 14 U-Net blocks in the encoder and decoder, yielding a model with 49.8 million parameters. We set the timestep dimensionality to 128 and train the model for 110,000 steps using the Adam optimizer with a linear warmup of 500 steps. The value of $\beta_1$ for the Adam optimizer is set to 0.5, and the learning rate is set to 0.0002. We also train an exponential moving average version of the U-Net. The audio is limited to a length of 8,387,584 samples (190 seconds), chosen so that the resulting spectrogram is divisible by the number of downsampling layers. We use a batch size of 4 and train the model on a single GPU with 16 GB of memory. The specific GPU used may vary depending on availability.

For the Neural Vocoder model, we set the width to 256 and use a single residual block between the input and output layers. This results in a model with 1.4 million parameters. We train the model for 40,000 steps using the Adam optimizer with a linear warmup of 500 steps. The value of $\beta_1$ for the Adam optimizer is set to 0.5, and the learning rate is set to 0.0002. The audio is limited to a length of 523,264 samples (11 seconds), chosen so that the resulting spectrogram is divisible by the number of downsampling layers. We use a batch size of 8 and train the model on a single GPU with 16 GB of memory. The specific GPU used may vary depending on availability.

Both of our models are trained using 16-bit floating point precision.

For diffusion, we use the diffusers library \cite{von-platen-etal-2022-diffusers} implementation of the DDIM algorithm \cite{song2020denoising} with 1000 timesteps and the cosine noise schedule from Glide \cite{nichol2021glide} (see Section \ref{appendix:hyperparameters} for more details).

\subsection{Evaluation}
Our model is currently under active development, so we have performed manual evaluation of the samples. In this evaluation, we have focused on the long-term coherence and harmony of the generated samples. We have randomly generated samples with different seeds, rather than cherry-picking specific samples. However, we have not yet implemented any quantitative metrics for evaluating the performance of the model.

\section{Results}
\label{sec:results}

\subsection{Sampling}
To evaluate the quality of the generated samples, we use subjective evaluations by human listeners. We generate our samples using 200 steps of DDIM and 200 iterations of GriffinLim.

Overall, we observe that the generated samples have good long-term coherence, with the ability to maintain coherence for approximately 3 minutes. The samples also display diverse structures, including repeating patterns throughout the entire song. However, the generated samples do exhibit some degradation in quality compared to human-generated music, particularly in terms of realism and naturalness. This may be due in part to the use of GriffinLim for phase reconstruction.

One particularly notable strength of the generated samples is their diversity, despite being trained on a relatively small dataset. This suggests that the model is able to learn generalizable patterns from the training data.

We also observe that the model struggles with global coherence early on in training, and that the loss does not show significant improvement as training progresses. However, we do notice that with longer training, the model is able to achieve better global coherence and overall sample quality improves.

We do not observe any significant improvements by increasing the number of DDIM sampling steps. This suggests that training the model with a shorter noise schedule may lead to faster sampling. Further experimentation will be necessary to confirm this possibility.
\begin{figure}[H]
    \centering
    \includegraphics[width=0.48\textwidth]{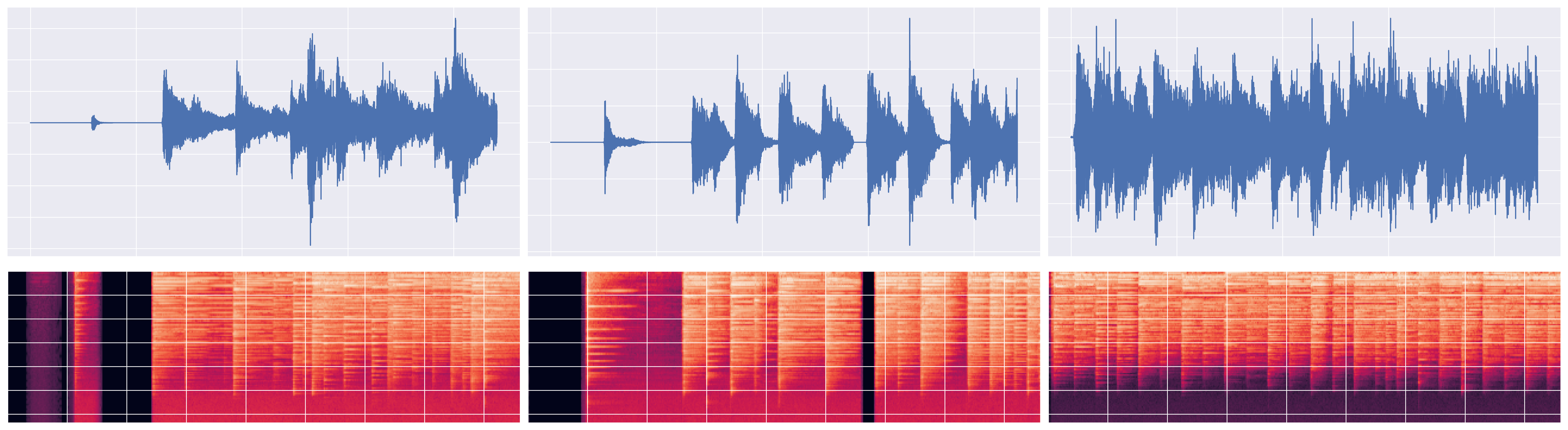}
    \caption{Illustrated from left to right are generated samples generated with 200 DDIM steps and 200 GriffinLim steps.}
    \label{fig:samplingresults}
\end{figure}

\subsection{Audio-to-Audio (Style Transfer)}
\label{ssec:audiotoaudio}

Starting with a mel spectrogram $\vx_0$, we use Equation (\ref{eqn:forwardnoisingprocess}) to add noise at a desired timestep $t$, resulting in a noised mel spectrogram $\vx_t$. We then employ the reverse diffusion process (described in Subsection \ref{sssec:reverseprocess}) to generate variations of the original audio that are more similar to our training data.

Our findings indicate that at low noise levels, the generated audio maintains the structure of the original audio while adapting the instruments and vocals to match those in the training data. However, the generated audio is too noisy at these low noise levels. On the other hand, when the noise level is higher, the generated audio more closely resembles the training data, but the structure of the original audio is largely lost.

We also note that percussive sounds, such as drums, are less sensitive to increases in noise levels. Even at high noise levels, the overall structure of the generated audio is preserved.

\begin{figure}[H]
    \centering
    \includegraphics[width=0.48\textwidth]{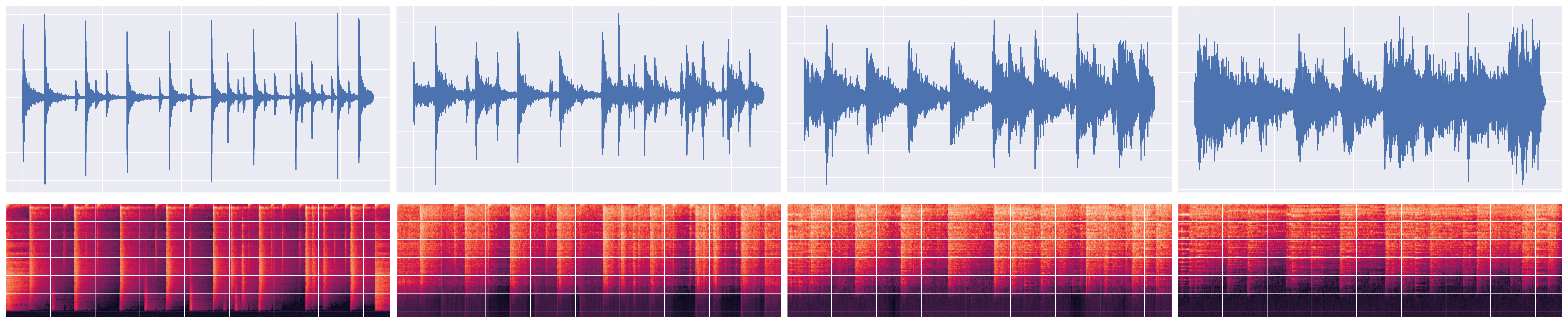}
    \caption{Illustrated, from left to right, are audio-to-audio samples of a drum loop with increasing noise timesteps, $t$: $t=0$, $t=100$, $t=500$, and $t=900$.}
    \label{fig:audio2audioresults}
\end{figure}

\subsection{Interpolation}
To perform audio interpolation using diffusion models, we start with two mel spectrograms $\vx_0^1$ and $\vx_0^2$ and use Equation (\ref{eqn:forwardnoisingprocess}) to add noise at the desired time step $t$, resulting in noised mel spectrograms $\vx_t^1$ and $\vx_t^2$. We then interpolate the two noised spectrograms as follows: $\vx_t = \gamma \vx_t^1 + (1 - \gamma) \vx_t^2$, where $\gamma \in \left[0, 1\right]$ is the interpolation ratio. We apply the reverse diffusion process (described in Subsection \ref{sssec:reverseprocess}) to generate interpolations of the two audio sources that are more similar to our training data.

We find that percussive sounds tend to be more prominent in the generated audio, even when the interpolation ratio is low. Similar to the results of the audio-to-audio task (see Subsection \ref{ssec:audiotoaudio}), we observe that at low noise levels the musical structure of the original audio is preserved, but the generated audio is noisy. On the other hand, at high noise levels, the musical structure of the generated audio closely resembles that of the training data.

\begin{figure}[H]
    \centering
    \includegraphics[width=0.48\textwidth]{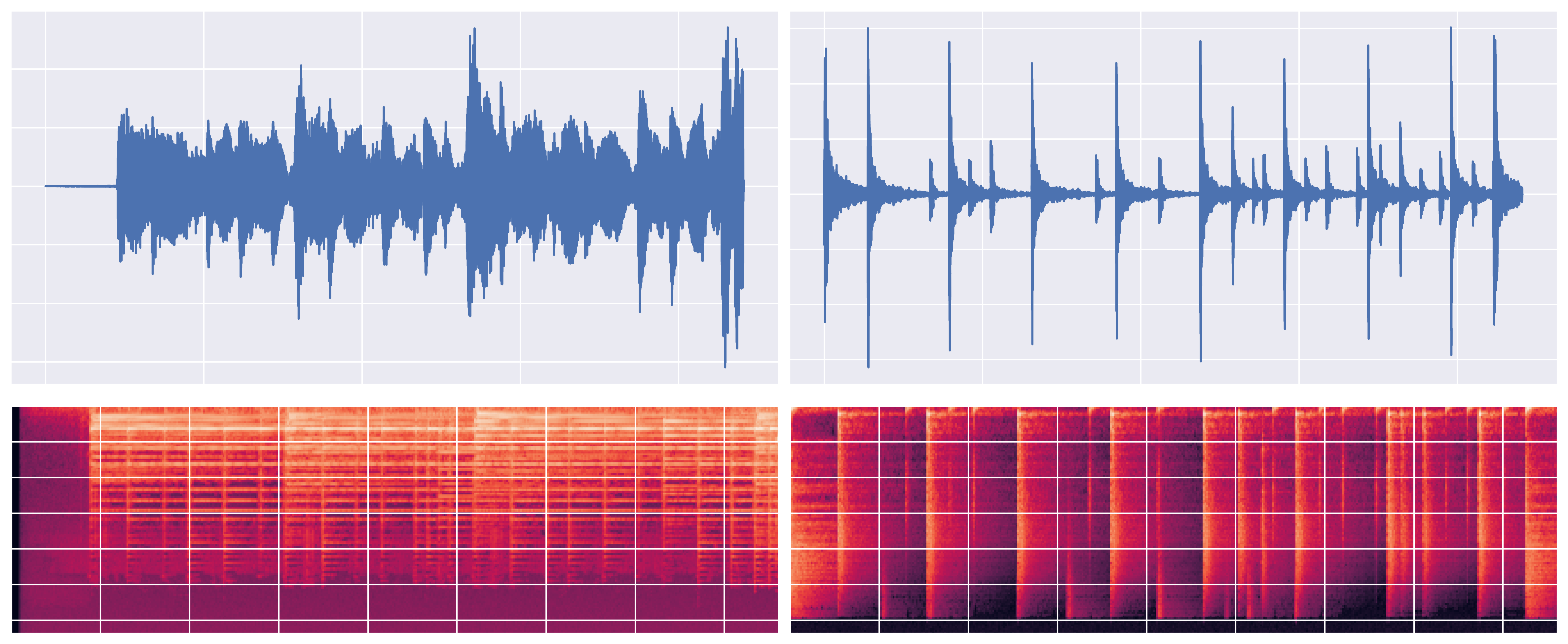}
    \caption{Original samples for interpolation, featuring a piano sample on the left and a drum loop on the right.}
    \label{fig:originalinterpolationsamples}
\end{figure}

\begin{figure}[H]
    \centering
    \includegraphics[width=0.48\textwidth]{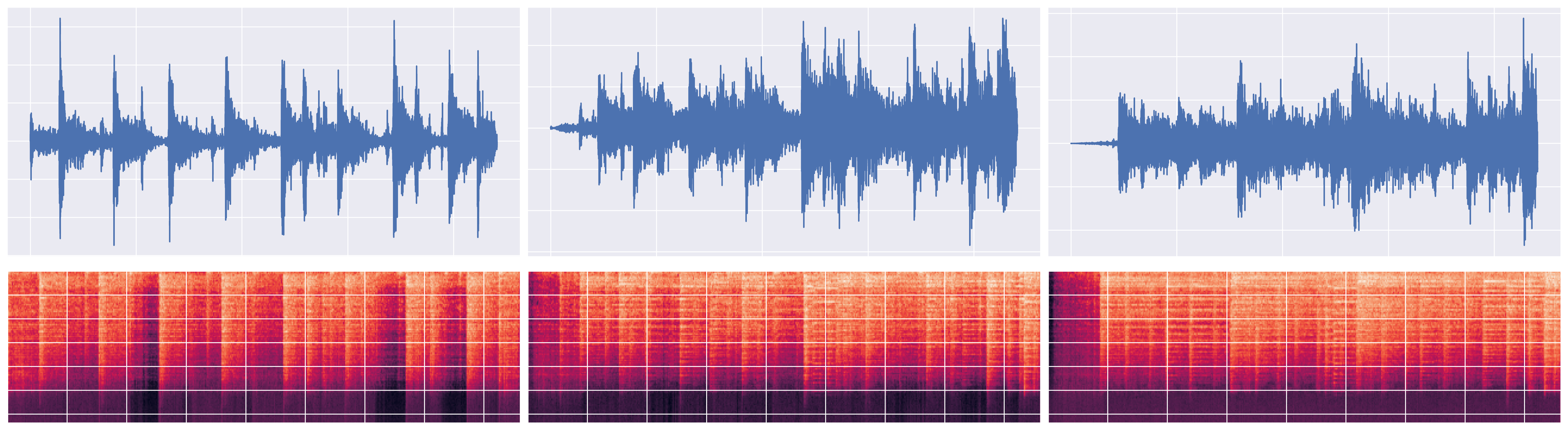}
    \caption{Illustrated, from left to right, are interpolation samples generated with a fixed noise timestep of $t=200$ and varying ratios of the original samples: 0.1, 0.5, and 0.9 representing the proportion of the first original sample in each interpolated sample, respectively}
    \label{fig:interpolationvaryingratioresults}
\end{figure}

\begin{figure}[H]
    \centering
    \includegraphics[width=0.48\textwidth]{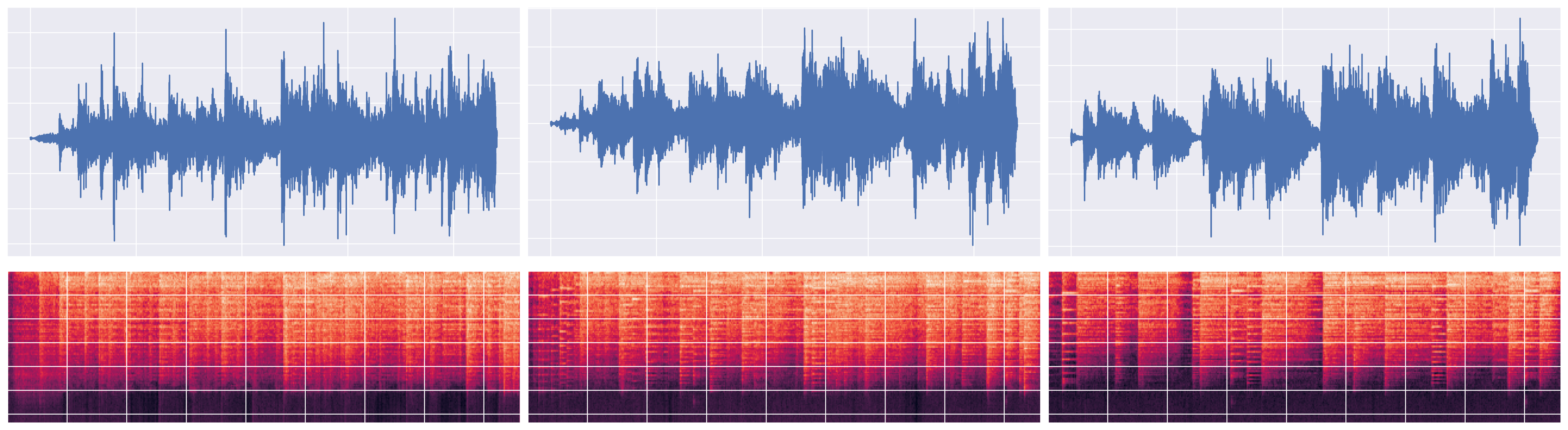}
    \caption{Illustrated, from left to right, are interpolation samples generated with a fixed ratio of 0.5 and increasing noise timesteps $t$: $t=100$, $t=500$, and $t=900$, respectively.}
    \label{fig:interpolationvaryinglevelresults}
\end{figure}

\subsection{Inpainting}
\label{ssec:inpainting}
In inpainting, we use a binary mask to identify the sections of an audio signal that should be kept and which should be removed. We then apply the Repaint algorithm \cite{lugmayr2022repaint}, as implemented by \cite{von-platen-etal-2022-diffusers}, to fill in the masked sections.

However, our results show that the inpainted sections often lack rhythm and do not accurately capture the melody and harmony of the original audio. The inpainted audio sounds like a completely novel sample and does not resemble the original audio at all. This leads to a sudden change in the musical structure of the song in the inpainted sections.

We are not sure why this issue occurs. Further experimentation will be necessary to investigate and address this problem.

\begin{figure}[H]
    \centering
    \includegraphics[width=0.48\textwidth]{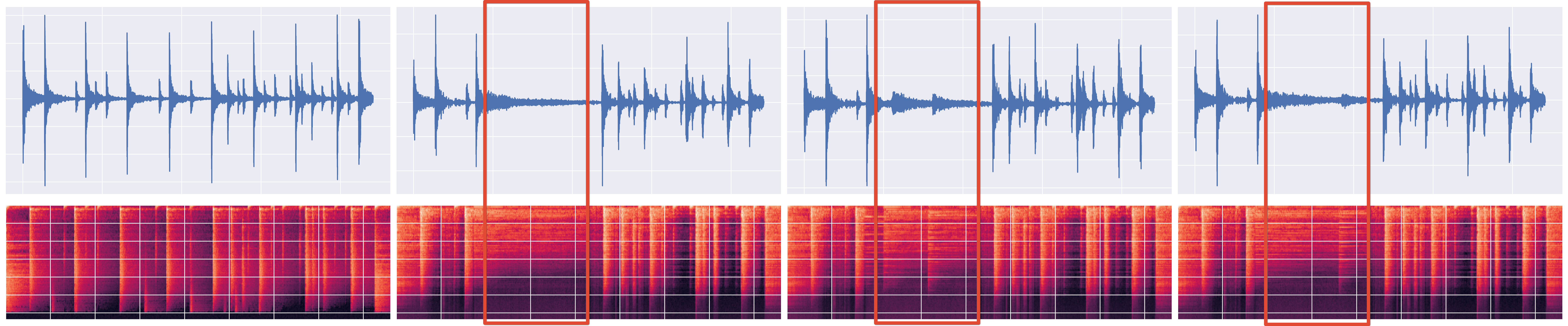}
    \caption{Illustrated from left to right: an original drum loop sample, and three inpainted versions generated using different seeds. The inpainted section in each sample is highlighted by the red box.}
    \label{fig:inpaintingresults}
\end{figure}

\subsection{Outpainting}
Outpainting involves extending the audio beyond the original recording by filling in additional sections with synthesized audio. To perform outpainting, we use the same algorithm as for inpainting (see Subsection \ref{ssec:inpainting}). We take half of the original audio and concatenate it with an empty spectrogram, then specify a mask to inpaint the empty part.

Unfortunately, our results are sub-optimal. The outpainted sections often sound different from the original audio, often lacking rhythm. This is more pronounced because the different spans of audio that are outpainted result in regions of sudden change, making the audio sound like multiple different sources concatenated together.

We are not sure why this issue occurs. Further experimentation will be necessary to investigate and address this problem.

\begin{figure}[H]
    \centering
    \includegraphics[width=0.48\textwidth]{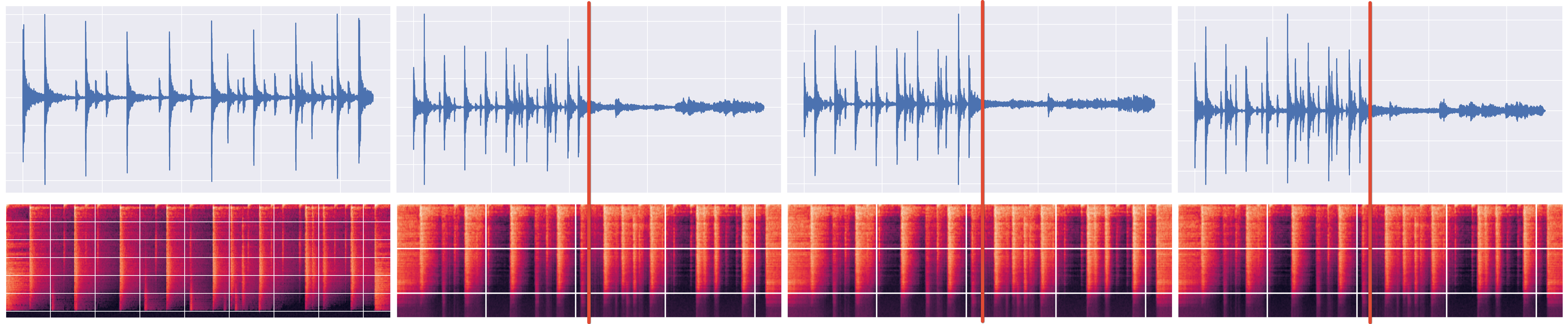}
    \caption{Illustrated from left to right: an original drum loop sample, and three outpainted versions generated using different seeds. The start of the outpainted section in each sample is highlighted by the red line.}
    \label{fig:outpaintingresults}
\end{figure}

\section{Future Work}
\label{sec:futurework}
There are several directions in which this work could be extended to improve the quality and usefulness of the approach in a music production setting. Some potential avenues for future research include:

\subsection{Conditional Generation}
In this work, we have focused on unconditional generation, meaning that there is little to no control over the synthesized samples. However, in a music production setting, it is important to allow some control over the generated music. Possible approaches for adding controllability to the synthesis process include:

\begin{itemize}
    \item Conditioning the model on factors such as lyrics, mood, or MIDI input. This could allow users to specify the content or style of the generated music.
    \item Allowing users to provide feedback during synthesis, either through explicit input or by interacting with the generated audio. This could allow users to shape the output in real-time or guide the synthesis process towards their desired results.
\end{itemize}

\subsection{Evaluation in Realistic Settings}
Our model has not yet been evaluated in a realistic setting, so it is important to determine its usefulness and effectiveness in music production. Possible approaches for evaluating the model include:

\begin{itemize}
    \item Conducting user studies or case studies to assess the usefulness and effectiveness of the techniques in a music production setting. This could involve gathering feedback from music producers, musicians, or other industry professionals.
    \item Comparing the model's output to that of human musicians or existing music production tools, using metrics such as subjective quality or task-specific performance. This could help to identify areas where the model excels or falls short compared to human or established standards.
\end{itemize}

\subsection{Generalization to Other Audio Tasks}
While this work has focused on music synthesis, we believe that the model could potentially be applied to other audio processing tasks as well. Some possible areas for exploration include:

\begin{itemize}
    \item Classification tasks, such as genre classification or instrument recognition. This could help to determine the model's ability to learn and represent musical concepts.
    \item Audio restoration tasks, such as noise reduction or audio enhancement. This could help to evaluate the model's ability to manipulate and improve audio signals.
\end{itemize}

\subsection{Improvements to Model Components}
There are several ways in which the performance of the model and its components could be improved:

\begin{itemize}
    \item Addressing the suboptimal performance of inpainting and outpainting for interactive audio design. This could involve redesigning the inpainting algorithm or incorporating new techniques for synthesizing audio.
    \item Scaling the model to improve its performance. This could help evaluate the scaling laws of the model.
    \item Incorporating new advancements in diffusion model sampling to allow for near real-time synthesis. This could help to make the model more responsive and interactive in a music production setting.
\end{itemize}

\subsection{Expanding the Range of Generated Music}
In this work, we have focused on generating music from a single instrument. However, there are many ways in which the model could be expanded to generate a wider range of music:

\begin{itemize}
    \item Examining how the model handles more complex data with multiple instruments and vocals. This could allow the model to generate a wider range of musical styles and textures.
    \item Exploring the use of multi-track or stem generation, allowing users to have explicit control over each generated instrument. This could be particularly useful in a music production setting.
    \item Investigating the use of the model for generating music in different styles or genres, such as electronic, classical, or world music. This could help to assess the model's ability to learn and represent diverse musical traditions.
\end{itemize}

Overall, there is significant potential for further research on our approach and its applications in the field of music synthesis AI, and we believe that these techniques have the potential to have a significant impact in the field.

\section{Conclusion}
\label{sec:conclusion}

In this work, we have introduced Msanii, a novel diffusion-based model for synthesizing long-context, high-fidelity music efficiently. By combining the expressiveness of mel spectrograms, the generative capabilities of diffusion models, and the vocoding capabilities of neural vocoders, Msanii is able to generate high quality audio. Our results demonstrate Msanii's capabilities for generating minutes of coherent audio efficiently, and we have also discussed the potential for Msanii to be used in various applications in the field of music production. With its strong performance and versatility, we believe that Msanii has significant potential for further research and development. Overall, Msanii shows promise as a powerful tool for music synthesis and production.

\section{Acknowledgement}
We would like to express our gratitude to OpenAI for providing access to ChatGPT, which has been instrumental in revising our paper. The use of ChatGPT has greatly improved the clarity and readability of our work, and we are grateful for the assistance it has provided.

\bibliographystyle{apalike}
\bibliography{msanii}

\clearpage
\appendix
\section{Data Preprocessing}
\label{appendix:datapreprocessing}

\subsection{Standard Scaling (Normalization)}
\begin{algorithm}[H]
    \caption{Standard Scaling Algorithm}
    \label{algo:standardscaling}
    \begin{algorithmic}[1]
        \Require{$x$: mini-Batch $B =\{x_{1 \ldots n}\}$, $m$: momentum, $d$: momentum decay, $\epsilon$: epsilon}
        \Ensure{$y = \frac{x - \mathbb{E}[x]}{\sqrt{\text{Var}[x] + \epsilon}}$}
        \Statex % empty line
        \If{no running statistics}
        \LineComment{Initialize running statistics}
        \State $\mu_R \gets \frac{1}{n} \sum_{i=1}^{n} x_i$ \Comment{mean}
        \State $\sigma^2_R \gets \frac{1}{n} \sum_{i=1}^{n} (x_i - \mu_{R})^2$ \Comment{variance}
        \EndIf
        \Statex % empty line
        \If{in training mode}
        \LineComment{Use mini-batch statistics}
        \State $\mu_{B} \gets \frac{1}{n} \sum_{i=1}^{n} x_i$ \Comment{mean}
        \State $\sigma^2_B \gets \frac{1}{n} \sum_{i=1}^{n} (x_i - \mu_{B})^2$ \Comment{variance}
        \Statex % empty line
        \LineComment{Update running statistics}
        \State $m \gets d \cdot m$ \Comment{momentum}
        \State $\mu_{R} \gets (1 - m) \cdot \mu_{R} + m \cdot \mu_{B}$ \Comment{mean}
        \State $\sigma^2_R \gets (1 - m) \cdot \sigma^2_R + m \cdot \sigma^2_B$ \Comment{variance}
        \Else
        \LineComment{Use running statistics}
        \State $\mu_{B} \gets \mu_R$ \Comment{mean}
        \State $\sigma^2_B \gets \mu_R$ \Comment{variance}
        \EndIf
        \Statex % empty line
        \State $y \gets \frac{x - \mu_B}{\sqrt{\sigma^2_B + \epsilon}}$
    \end{algorithmic}
\end{algorithm}
This algorithm standardizes the elements of the input $x$ using the mean $\mathbb{E}[x]$ and variance $\text{Var}[x]$ of the input. The momentum $m$ and momentum decay $d$ are used to update the running statistics during training, and the small constant $\epsilon$ is used to prevent division by zero. If the model is in training mode, the batch statistics $\mu_{B}$ and $\sigma^2_B$ are used to compute the mean and variance. If the model is in inference mode, the running statistics $\mu_{R}$ and $\sigma^2_R$ are used instead.
\vfill % Prevents content from being spaced out to fill the remaining space

\subsection{Min-Max Scaling}
\begin{algorithm}[H]
    \caption{Min-Max Scaling Algorithm}
    \label{algo:minmaxscaling}
    \begin{algorithmic}[1]
        \Require{$x$: mini-Batch $B =\{x_{1 \ldots n}\}$, $y_{min}$: minimum value after scaling, $y_{max}$: maximum value after scaling,  $m$: momentum, $d$: momentum decay}
        \Ensure{$y = \frac{x - \text{min}(x)}{\text{max}(x) - \text{min}(x)} * (y_{max} - y_{min}) + y_{min}$}
        \Statex % empty line
        \If{no running statistics}
        \LineComment{Initialize running statistics}
        \State $\text{min}_R \gets \text{min}(x)$ \Comment{min}
        \State $\text{max}_R\gets \text{max}(x)$ \Comment{max}
        \EndIf
        \Statex % empty line
        \If{in training mode}
        \LineComment{Use mini-batch statistics}
        \State $\text{min}_B \gets \text{min}(x)$ \Comment{min}
        \State $\text{max}_B \gets \text{max}(x)$ \Comment{max}
        \Statex % empty line
        \LineComment{Update running statistics}
        \State $m \gets d \cdot m$ \Comment{momentum}
        \State $\text{min}_{R} \gets (1 - m) \cdot \text{min}_{R} + m \cdot \text{min}_{B}$ \Comment{min}
        \State $\text{max}_{R} \gets (1 - m) \cdot \text{max}_{R} + m \cdot \text{max}_{B}$ \Comment{max}
        \Else
        \LineComment{Use running statistics}
        \State $\text{min}_{B} \gets \text{min}_R$
        \State $\text{max}_B \gets \text{max}_R$
        \EndIf
        \Statex % empty line
        \State $y = \frac{x - \text{min}_B}{\text{max}_B - \text{min}_B} * (y_{max} - y_{min}) + y_{min}$
        \State $y = \text{max}(y_{min}, \text{min}(y, y_{max}))$ \Comment{clamp $y$ to be in the range $\left[y_{min}, y_{max}\right]$}
    \end{algorithmic}
\end{algorithm}
This algorithm scales the elements of the input $x$ to be in the range $\left[y_{min}, y_{max}\right]$. The momentum $m$ and momentum decay $d$ are used to update the running statistics during training. If the model is in training mode, the batch statistics $\text{min}_{B}$ and $\text{max}_B$ are used to compute the min and max. If the model is in inference mode, the running statistics $\text{min}_{R}$ and $\text{max}_R$ are used instead.

\clearpage
\section{Hyper-parameters}
\label{appendix:hyperparameters}

\subsection{Neural Vocoder}
\begin{table}[htb!]
    \centering
    \begin{tabular}{l|c}
        \toprule
        \textbf{Hyper-parameter}          & \textbf{Value} \\
        \midrule
        Data                              &                \\
        \quad Sample rate                 & 44100          \\
        \quad Audio length                & 523264         \\
        \quad Audio channels              & 2              \\
        \quad Batch size                  & 8              \\
                                          &                \\
        Transforms                        &                \\
        \quad FFT size                    & 2048           \\
        \quad Window length               & 2048           \\
        \quad Hop length                  & 1024           \\
        \quad Mel frequencies             & 128            \\
        \quad Feature scaling momentum    & 0.001          \\
        \quad Feature scaling decay       & 0.99           \\
        \quad GriffinLim iterations       & 200            \\
                                          &                \\
        Vocoder                           &                \\
        \quad Model dimension             & 256            \\
        \quad MLP hidden dimension factor & 4              \\
                                          &                \\
        Training                          &                \\
        \quad Learning rate (lr)          & 0.0002         \\
        \quad Optimizer                   & Adam           \\
        \quad Adam betas                  & 0.5, 0.999     \\
        \quad lr warmup iterations        & 500            \\
        \quad lr warmup start factor      & $\frac{1}{3}$  \\
        \quad Precision                   & 16             \\
        \quad Training steps              & 40000          \\
        \bottomrule
    \end{tabular}
    \caption{Neural Vocoder hyper-parameters.}
\end{table}
\vfill % Prevents content from being spaced out to fill the remaining space

\subsection{U-Net}
\begin{table}[H]
    \centering
    \begin{tabular}{l|c}
        \toprule
        \textbf{Hyper-parameter}           & \textbf{Value}                  \\
        \midrule
        Data                               &                                 \\
        \quad Sample rate                  & 44100                           \\
        \quad Audio Length                 & 8387584                         \\
        \quad Audio channels               & 2                               \\
        \quad Batch size                   & 4                               \\
                                           &                                 \\
        Diffusion                          &                                 \\
        \quad Noise schedule               & cosine                          \\
        \quad Number of training timesteps & 1000                            \\
        \quad Number of sampling steps     & 200                             \\
                                           &                                 \\
        U-Net                              &                                 \\
        \quad Base model dimension         & 256                             \\
        \quad Timestep dimension           & 128                             \\
        \quad MLP hidden dimension factor  & 4                               \\
        \quad Number of attention heads    & 8                               \\
        \quad Dimensionality factor        & 1,1,1,1,1,1,1                   \\
        \quad Dilations                    & 1,1,1,1,1,1,1                   \\
        \quad Has attention                & \uppercase{f, f, f, f, f, t, t} \\
        \quad Has resampling               & \uppercase{t, t, t, t, t, t, f} \\
        \quad Blocks per resolution        & 2,2,2,2,2,2,2                   \\
                                           &                                 \\
        EMA U-Net                          &                                 \\
        \quad Start step                   & 2000                            \\
        \quad Decay                        & 0.995                           \\
        \quad Update every n-steps         & 10                              \\
                                           &                                 \\
        Training                           &                                 \\
        \quad Learning rate (lr)           & 0.0002                          \\
        \quad Optimizer                    & Adam                            \\
        \quad Adam betas                   & 0.5, 0.999                      \\
        \quad lr warmup iterations         & 500                             \\
        \quad lr warmup start factor       & $\frac{1}{3}$                   \\
        \quad Precision                    & 16                              \\
        \quad Training steps               & 110000                          \\
        \bottomrule
    \end{tabular}
    \caption{U-Net hyper-parameters.}
\end{table}

\clearpage
\section{Optimizing U-Net Width for Improved Performance}
The U-Net architecture, similar to transformers, can achieve significant performance gains by increasing its width. However, it is essential to consider the relationship between the width of the U-Net and the dimension of the spectrogram frequencies when using it as a feature.

For instance, using a U-Net with a width that is lower than the frequency dimension of the spectrogram forces the input layer to only learn the principal components of the mel spectrogram. This approach may work well for a clean mel spectrogram but is not suitable when the mel spectrogram is corrupted with noise.

Our experimentation has shown that it is crucial for the U-Net width to be at least 2x larger than the frequency dimension of the spectrogram to ensure optimal performance. Further experimentation is required to fully understand this phenomenon and to optimize the U-Net's width for improved performance.

\end{document}